\documentclass[aps,twocolumn,preprintnumbers,groupedaddress,superscriptaddress,floatfix,reprint,nofootinbib,longbibliography]{revtex4-1}
\usepackage{mathrsfs}
\usepackage{natbib}
\usepackage{verbatim}
\usepackage{enumerate}
\usepackage{graphicx,bbm,amsbsy,amsfonts,amssymb,amsmath}
\usepackage{bm}
\usepackage{hyperref}
\usepackage{booktabs}
\usepackage{gensymb}
\usepackage{xcolor}

\usepackage{slashed}

\usepackage{footmisc}

\usepackage{blindtext}

\hypersetup{
    colorlinks=true,       
    linkcolor=red,          
    citecolor=blue,        
    filecolor=magenta,      
    urlcolor=blue           
}

\def\tr{\text{tr}\,}

\def\ZZ{\mathbb{Z}} 
\def\RR{\mathbb{R}}

\def\tr{\text{tr}}

\def\2{{(2)}}
\def\1{{(1)}}
\def\0{{(0)}}
\def\m1{{(-1)}}

\let\vec=\mathbf

\let\oldAA\AA
\renewcommand{\AA}{\text{\normalfont\oldAA}}

\begin{document}

\preprint{UMN--TH--4403/24, FTPI--MINN--24/23 }

\vspace*{1mm}

\title{Cheshire \texorpdfstring{$\theta$}{theta} terms, Aharonov-Bohm effects, and axions}

\author{Shi Chen}
\email{chen8743@umn.edu}
\affiliation{School of Physics and Astronomy, 
University of Minnesota, Minneapolis MN 55455, USA}
\author{Aleksey Cherman}
\email{acherman@umn.edu}
\affiliation{School of Physics and Astronomy, 
University of Minnesota, Minneapolis MN 55455, USA}
\author{Gongjun Choi}
 \email{choi0988@umn.edu}
 \affiliation{William I. Fine Theoretical Physics Institute, School of Physics and Astronomy,
University of Minnesota, Minneapolis, MN 55455, USA}
\author{Maria Neuzil}
 \email{neuzi008@umn.edu}
 \affiliation{School of Physics and Astronomy, 
University of Minnesota, Minneapolis MN 55455, USA}

\begin{abstract}
We discuss unusual $\theta$ terms that can appear in field theories that 
allow global vortices.
These `Cheshire $\theta$ terms' induce Aharonov-Bohm effects for some particles that move around vortices. 
For example, a Cheshire $\theta$ term can appear in QCD coupled to an axion and induces Aharonov-Bohm effects for baryons and leptons moving around axion strings. 
\end{abstract}

\maketitle
\flushbottom


{\bf Introduction.}  
Periodic couplings are common in applications of quantum field theory. 
Quantum chromodynamics (QCD) has a $2\pi$-periodic $\theta$ parameter~\cite{Polyakov:1975rs,tHooft:1976rip,Callan:1976je,Jackiw:1976pf}, as do many effective field theory (EFT) descriptions of condensed matter systems, see e.g.~Refs.~\cite{Haldane:1982rj,Haldane:1983ru,Haldane:1988zz,Fradkin:1986pq,Wilczek:1987mv,Abanov:1999qz,Ryu:2010ah}.   
These parameters appear in Euclidean actions $S$ in $D$ spacetime dimensions via $i \theta \int d^{D}x\, q(x)$, where $\int d^{D}x\,q(x) \in \mathbb{Z}$ is the instanton topological charge of some field. 
Physics depends on $S$ via $e^{-S}$, leading to $\theta \simeq \theta +2\pi$. 

$\theta$ terms have many effects.  
For example, spin-$1/2$ antiferromagnetic spin chains in one spatial dimension, which have an EFT description via an $O(3)$ sigma model with $\theta = \pi$, are gapless, while spin-$1$ spin chains, which correspond to $\theta = 2\pi$, are gapped~\cite{Haldane:1982rj,Haldane:1983ru,Haldane:1988zz, Zamolodchikov:1992zr}.
In particle physics, QCD is viewed as an EFT for the Standard Model (SM) and any beyond-SM physics.  
P and CP symmetries are violated in the SM, so it is natural to expect the P- and CP-symmetry-violating QCD $\theta$ parameter to be $\mathcal{O}(1)$.  
This would lead to an electric dipole moment for the neutron $d_n\sim 6 \times 10^{-17}\textrm{e}\cdot\mathrm{cm}$, see e.g.~Refs.~\cite{Pospelov:2005pr,Engel:2013lsa,Chupp:2017rkp}.  
The famous `strong CP problem' is to understand why the experimental value of $d_n$ is at least $10^{-10}$ times smaller~\cite{Abel:2020pzs,ParticleDataGroup:2024cfk}.

Here we study another type of $\theta$ parameter, denoted by $\hat{\theta}$, associated with what we call a `Cheshire $\hat{\theta}$ term.' 
Cheshire $\hat{\theta}$ terms can appear in field theories with a $2\pi$-periodic compact scalar field $\varphi$ (such as a $U(1)$ pseudo-Nambu-Goldstone boson) together with another $0$-form Abelian global symmetry $G$. 
Such theories necessarily have a $(D-2)$-form $U(1)$ winding symmetry~\cite{Gaiotto:2014kfa,Delacretaz:2019brr} which counts vortices of $\varphi$. 
When $G =U(1)$ with a conserved current $j$, the Cheshire $\hat{\theta}$ term is given by
\begin{align} 
i\hat{\theta} \int d^{D}x \frac{\partial_{\mu}\varphi}{2\pi} j^{\mu}
\label{eq:new_theta_generic}
\end{align}
in the leading order.
Naively, this term seems physically invisible since it vanishes after integrating by parts and using $\partial_{\mu} j^{\mu} = 0$. 
Such terms are usually excluded from EFT actions, see e.g.~Refs.~\cite{Georgi:1991ch,Henning:2015daa}.  
Indeed, the particle spectrum and particle-particle scattering amplitudes do not depend on $\hat{\theta}$.
However, $\hat{\theta}$ does influence some correlation functions of point operators.  Most importantly, it leads to an Aharonov-Bohm (AB) effect characterized by the phase $e^{i\hat{\theta}}$ for particles with unit $U(1)$-charge moving around unit-winding $\varphi$ vortices. 
We explain these effects in the subsequent sections, and then describe an application to the SM coupled to an axion.



%
\vspace{0.5em}
{\bf General idea.}\footnote{
We use differential form notation here, but return to index notation when discussing axions in $D=4$.} 
Consider a field theory in a $D$-dimensional spacetime, with $D>1$, that satisfies the the following three assumptions.
(I) One of the dynamical fields is a $2\pi$-periodic scalar $\varphi(x)$, and its conserved winding number gives rise to a global $(D\!-\!2)$-form symmetry $U(1)_1^{[D\!-\!2]}$.
(II) There is also another global 0-form symmetry $U(1)_2^{[0]}$.
(III) The $U(1)_1^{[D\!-\!2]}$ and $U(1)_2^{[0]}$ symmetries do not have individual or mixed 't Hooft anomalies with each other.
Then, in the presence of a $U(1)_2^{[0]}$ background gauge field $A_2$, the partition function of this theory can be expressed by the path integral
\begin{align}
    \int\mathcal{D}\varphi\, \exp\Big[ -S_{\varphi}(\varphi,A_2)\Big]\,.
    \label{eq:formal_partition_function}
\end{align}
To make Eq.~\eqref{eq:formal_partition_function} simple, we have formally integrated out dynamical fields other than $\varphi$ to obtain $S_{\varphi}(\varphi,A_2)$.
Our assumptions imply that $\exp[ -S_{\varphi}(\varphi,A_2+\mathrm{d}\chi)]=\exp[ -S_{\varphi}(\varphi,A_2)]$ 
for any $2\pi$-periodic scalar $\chi(x)$.

Now we define a new family of theories parameterized by a real coupling constant $\hat{\theta}$, with the partition function
\begin{align}
    \mathcal{Z}_{\hat{\theta}} = \int\mathcal{D}\varphi\ \exp\left[ -S_{\varphi}\left(\varphi,\frac{\hat{\theta}}{2\pi}\mathrm{d}\varphi\right)\right]\,.
    \label{eq:Ztheta}
\end{align}
The discussion above implies that $\hat{\theta}$ is $2\pi$-periodic, i.e.
\begin{align}
    \mathcal{Z}_{\hat{\theta}} = \mathcal{Z}_{\hat{\theta}+2\pi}\,.
\end{align}
This construction is a general form of the Cheshire $\hat{\theta}$ term in Eq.~\eqref{eq:new_theta_generic}. The Cheshire $\hat{\theta}$ term can be thought of as a special kind of gauging of $U(1)_2^{[0]}$ with a gauge field $a_2 = \frac{\hat{\theta}}{2\pi}\mathrm{d}\varphi$.  Since the path integral sums over $\varphi$, it also sums over $a_2$. In this sense $a_2$ is a ``dynamical'' gauge field, which we emphasize by writing it in lower case.   However, $a_2$ is flat, $d a_2 = 0$, so passing from Eq.~\eqref{eq:formal_partition_function} to Eq.~\eqref{eq:Ztheta} does not introduce any new propagating modes.  Moreover, unlike a typical example of flat gauging where one does an unconstrained sum over flat gauge fields, here $a_2$ is fixed in terms of $\varphi$ up to gauge transformations. 
Therefore, if we evaluate the partition function on some $S^D$ (an infrared-regulated version of $\mathbb{R}^D$), we can always rotate $\hat{\theta}$ back to zero, i.e.
\begin{equation}
    \mathcal{Z}_{\hat{\theta}}(S^D) = \mathcal{Z}_{0}(S^D)\,,
\end{equation}
by a gauge transformation.
This means that $\hat{\theta}$ is invisible in the local physics: the particle spectrum, the scattering between particles, etc.~are not affected by $\hat{\theta}$.

The Cheshire $\hat{\theta}$ term does lead to visible physical consequences as long as the spacetime topology is more complicated.
This can be achieved either by considering ``thermodynamic'' spacetime manifolds $M^{D-1}\times S^1$ or by inserting $\varphi$ vortices.
Vortex operators $\mathcal{V}_n$ with charge $n$ under $U(1)_1^{[D-2]}$ induce a locally-flat $U(1)_2^{[0]}$ gauge field $a_2 = \frac{\hat{\theta}}{2\pi}\mathrm{d}\varphi$ with a non-trivial holonomy, while point operators $\mathcal{O}_q$ with charge $q$ under $U(1)_2^{[0]}$ induce a locally-flat background gauge field $A_1$ for the $U(1)_1^{[D-2]}$ symmetry:
\begin{subequations}
\label{eq:holonomies}
 \begin{align}
    e^{i \int_C a_2} &= \mathrm{e}^{i\hat{\theta}n}\\
    e^{i \int_{S^{D-1}} A_1} &= \mathrm{e}^{i (-1)^{D-1}\hat{\theta}q} \,,
\end{align}   
\end{subequations}
assuming that $C$ has unit linking with the vortex worldvolume $S$ while $S^{D-1}$ has unit linking with the location of $\mathcal{O}_q(x)$. 
When $D>2$, this implies that winding-$n$ $\varphi$ vortices induce an Aharonov-Bohm (AB) effect for particles with $U(1)_2^{[0]}$ charge $q$: a charge-$q$ particle traveling along $C$ picks up a phase shift of
\begin{align}
e^{i q n \hat{\theta}} \,.
\label{eq:AB_phase}
\end{align}
We will refer to this as `picking up an AB phase' below.  For any given $\hat{\theta}$, $\mathcal{O}_q(x)$ and $\mathcal{V}_n(S)$ are both in the vacuum sector, so we emphasize that Eq.~\eqref{eq:AB_phase} is \emph{not} a braiding phase for point and vortex operators. 
Such braiding phases could only appear between operators in distinct twisted sectors.

These phenomena can be interpreted as a Cheshire version of the Witten effect~\cite{Witten:1979ey}. 
Indeed, if our theory had a $U(1)_1^{[0]}$  shift symmetry for $\varphi$, we would see a conventional Witten effect: the charge-$q$ operators under $U(1)^{[0]}_2$ have a fractional   $U(1)_1^{[0]}$ charge $q\hat{\theta}/2\pi$. But if   $U(1)_1^{[0]}$ is explicitly broken, its charge becomes ill-defined and the Witten effect disappears.
However, the holonomies and AB effects in Eqs.~\eqref{eq:holonomies}, \eqref{eq:AB_phase} are vastly more robust. 
In this sense, they are better regarded as the very essence of the Witten effect; fractional charges are just bonuses caused by extra symmetries with extra mixed 't Hooft anomalies.

Some of the assumptions we made at the beginning of this section can be relaxed.\footnote{
In fact, there is a vastly generalized framework for describing $\theta$ angles which only requires the existence of a non-anomalous symmetry~\cite{ChenChermanNeuzil}, with the electromagnetism $\theta$ angle and Cheshire $\hat{\theta}$ terms as its simplest examples.}
For example, we can reduce $U(1)^{[0]}_2$ to a discrete symmetry $(\mathbb{Z}_{k})^{[0]}_2$.
A background gauge field for $\mathbb{Z}_{k}$ can be described by a $U(1)$ gauge field $A$ together with the constraint $kA=\mathrm{d} C$
where $C$ is a $2\pi$-periodic scalar.
Therefore, we can still introduce the $2\pi$-periodic Cheshire $\hat{\theta}$ coupling as long as it is quantized properly, $\hat{\theta} \in \frac{2\pi}{k}\mathbb{Z}$.
Another way that $\hat{\theta}$ can get quantized is if there is a global symmetry that flips the sign of the Cheshire $\hat{\theta}$ term, in which case the symmetry-preserving values of $\hat{\theta}$ are $0$ and $\pi$.   

\vspace{0.5em}
{\bf 2D model.} 
We now illustrate the discussion in a simple family of models $\mathcal{T}_{\hat{\theta}}$ in $D=2$ involving two $2\pi$-periodic scalar fields $\varphi_1, \varphi_2$ with the Euclidean action
\begin{align}\label{eq:two_scalar}
   \int\!\left[\sum_{i=1,2}\! \!\frac{R_i^2}{4\pi} ||\mathrm{d}\varphi_i||^2 
    - \frac{i\hat{\theta}}{4\pi^2} \mathrm{d}\varphi_1\wedge \mathrm{d}\varphi_2 + \star \mathcal{V}(\varphi_1,\varphi_2)\right]
\end{align}
where $||d\varphi_i||^2 = d\varphi_{i} \wedge \star d\varphi_i$, $R_1,R_2$ are dimensionless constants, and $\mathcal{V}(\varphi_1,\varphi_2)$ is a potential on the target space $T^2$.
The conserved winding numbers of $\varphi_1, \varphi_2$ give rise to two $U(1)^{[0]}$ symmetries, $U(1)^{[0]}_1$ and $U(1)^{[0]}_2$, with charges $m_1,m_2 \in \ZZ$. 
The current associated with e.g. $U(1)_2^{[0]}$ is $j_2 = (2\pi)^{-1} \star d\varphi_2$, so the $\hat{\theta}$ term in Eq.~\eqref{eq:two_scalar} is equivalent to $i\hat{\theta} \int \frac{d\varphi_1}{2\pi} \wedge \star j_2$, matching Eq.~\eqref{eq:new_theta_generic} with $D=2$.

When $\mathcal{V}(\varphi_1,\varphi_2) = 0$, there are two extra $U(1)^{[0]}$ symmetries associated with shifts of $\varphi_1, \varphi_2$ with charges $e_1,e_2 \in \ZZ$.  
Then $\mathcal{T}_{\hat{\theta}}$ is a non-chiral $c=2$ conformal field theory with the chiral algebra $\hat{\mathfrak{u}}(1)_+^2\!\times\! \hat{\mathfrak{u}}(1)_-^2$.
The $\hat{\mathfrak{u}}(1)_+^2\!\times\! \hat{\mathfrak{u}}(1)_-^2$ primaries $\mathcal{O}_{\hat{\theta}}\left(\substack{m_1,e_1\\m_2,e_2}\right)$ are labeled by four charges and have the conformal dimensions
\small
\begin{equation*}
    h_{\pm}\! = \frac{1}{4}\!\left(m_1R_1 \pm \frac{e_1+\frac{\hat{\theta}}{2\pi}m_2}{R_1} \right)^2
    \!\!\!+\frac{1}{4}\! \left(m_2R_2 \pm \frac{e_2-\frac{\hat{\theta}}{2\pi}m_1}{R_2} \right)^2\!\! .
\end{equation*}\normalsize
For any $\hat{\theta}$, primaries are local when $m_1,e_1,m_2,e_2\in\ZZ$.
Local primaries always have integral spins, $h_+-h_-\in\mathbb{Z}$, and their correlation functions are always single-valued without any branch cut thanks to a 2-dimensional Dirac-Schwinger-Zwanziger quantization condition.
This model features a 2-dimensional Witten effect, meaning a correspondence between $\mathcal{T}_{\hat{\theta}}$ primaries and $\mathcal{T}_0$ primaries:
\begin{equation}
    \mathcal{O}_{\hat{\theta}}\left(\substack{m_1,e_1\\m_2,e_2}\right) \quad\leftrightarrow\quad \mathcal{O}_{0}\left(\substack{m_1,e_1+\hat{\theta}m_2/2\pi\\m_2,e_2-\hat{\theta}m_1/2\pi}\right)\,.
\end{equation}
For a generic $\hat{\theta}$, since the induced charges are fractional, local primaries on one side of this correspondence map to non-local primaries that live in $U(1)^{[0]}_1\times U(1)^{[0]}_2$-twisted sectors on the other side of the correspondence.  

Next, suppose that $\mathcal{V}(\varphi_1,\varphi_2)\neq0$.  Then  $e_1,e_2$ cease to be good quantum numbers.
Nevertheless, an important part of the Witten effect survives because $m_1$ and $m_2$ remain conserved quantum numbers.
Namely, a charge-$(m_1,m_2)$ local operator in $\mathcal{T}_{\hat{\theta}}$ corresponds to a charge-$(m_1,m_2)$ non-local operator in $\mathcal{T}_0$ which lives in a $U(1)^{[0]}_1\times U(1)^{[0]}_2$-twisted sector labeled by $(\mathrm{e}^{\mathrm{i}\hat{\theta}m_2},\mathrm{e}^{-\mathrm{i}\hat{\theta}m_1})$.  This can be summarized as
\begin{align}
    \mathcal{O}_{\hat{\theta}}\!\left(\substack{m_1
    \\m_2}\right)
    \ \leftrightarrow\ 
    \mathcal{O}_{0}\!\left(\substack{m_1
    \\m_2}\right) \!\text{ twisted by }\! \substack{\exp\left(\!\mathrm{i}\hat{\theta}m_2\!\right)\\\exp\left(\!-\mathrm{i}\hat{\theta}m_1\!\right)}\in\substack{U(1)^{[0]}_1\\U(1)^{[0]}_2}.
\end{align}
    We view this reshuffle of twisted sectors as the essence of the Witten effect, and refer to it as the `primary Witten effect'. The fractional charges that appear when $\mathcal{V} = 0$, which is what is usually viewed as the definition of the Witten effect, are then relegated to a secondary role. We thus refer to charge fractionalization as the `secondary Witten effect'.  To obtain the secondary Witten effect one needs two ingredients.  First, one needs the primary Witten effect that reshuffles the $U(1)^{[0]}_1\times U(1)^{[0]}_2$-twisted sectors.  Second, one needs the presence of extra symmetries (the shift symmetries at $\mathcal{V} = 0$) that have appropriate\footnote{
Suppose a $U(1)^{[p]}$ symmetry and a $U(1)^{[D-p-2]}$ symmetry have a mixed anomaly with the inflow action $\int A^{[p+1]}\wedge\mathrm{d}B^{[D-p-1]}$.
Then the twisted sector with the $U(1)^{[D-p-2]}$-twist $\mathrm{e}^{\mathrm{i}\beta}$ necessarily carries a fractional $U(1)^{[p]}$-charge ${\beta}/{2\pi}\mod 1$, and vice versa.
} 
mixed 't Hooft anomalies with $U(1)^{[0]}_1\times U(1)^{[0]}_2$.

\vspace{0.5em}
{\bf Axion-QCD EFT.} We now discuss a Cheshire $\hat{\theta}$ term in high energy particle physics.  Between the weak scale $m_W$ and the QCD strong scale $\Lambda_{\rm QCD}$, the interactions of quarks, photons, and
leptons can be described by a QCD+QED EFT.  We neglect QED below since it does not affect our discussion.  One approach to the strong CP problem of this EFT is to couple QCD
to the Peccei-Quinn (PQ) axion
$\varphi$~\cite{Peccei:1977ur,Peccei:1977hh,Weinberg:1977ma,Wilczek:1977pj}, a $2\pi$-periodic NG boson for a spontaneously-broken $U(1)_{\rm PQ}$ symmetry,
\begin{align}
    &\frac{i}{32\pi^2} \int d^{4}x\, \varphi\, \tr f_{\mu\nu} \tilde{f}^{\mu\nu} + \int d^{4}x\, \frac{f_{\rm PQ}^2}{2} (\partial_{\mu}\varphi)^2
    \,,
    \label{eq:axion}    
\end{align}
where $f_{\mu\nu}$ is the field strength of the $SU(N = 3)$ color gauge field with Lie algebra generators $\tr \, t_i t_j = \frac{1}{2}\delta_{ij}$, and $\tilde{f}_{\mu\nu} = \epsilon_{\mu\nu\alpha\beta}f^{\alpha\beta}$.
The QCD
$\theta$ term can be removed by shifting $\varphi$.  If $U(1)_{\rm PQ}$ is explicitly broken by only the QCD ABJ anomaly,\footnote{Ensuring that this is true to a sufficiently good approximation is known as the axion quality problem. Its solution requires considerable effort, see e.g.
Ref.~\cite{Witten:1984dg,Kim:1984pt,Kamionkowski:1992mf,Barr:1992qq,Randall:1992ut,Ghigna:1992iv,Holman:1992us,Dine:1992vx,Choi:2003wr,Svrcek:2006yi}.} the minimum of the effective axion potential is believed to be very close to zero, resolving the strong CP problem. Coupling QCD to the axion is thus a way of getting rid of a CP-violating
periodic coupling.

Amusingly, the axion-SM EFT actually has at least one more periodic coupling.  If the neutrinos were massless, the SM would have a $U(1)_{B-L}$ symmetry.  If the observed neutrino masses arise from Dirac mass terms involving right-handed neutrinos, then $U(1)_{B-L}$ survives even when the neutrinos are massive, and is free of 't Hooft anomalies. In this situation we can write a  Cheshire $\hat{\theta}$ term 
\begin{align}
    i \hat{\theta} \int d^{4}x\, 
    \frac{\partial_{\mu}\varphi}{2\pi} j^{\mu}_{B-L}\,,
    \label{eq:theta_hat_QCD}
\end{align}
where $j_{B-L}$ is the difference of the baryon number $j_B$ and lepton number $j_L$
currents.\footnote{A coupling of this basic form was written down in Ref.~\cite{Nagasawa:1997zn}, but the periodicity of $\hat{\theta}$ and the fact that $j$ in Eq.~\eqref{eq:theta_hat_QCD} must be conserved were not discussed there.}   The other combination of $U(1)_B$ and $U(1)_L$, $U_{B+L}$, is broken to $(\ZZ_3)_{B+L}$ by an Adler-Bell-Jackiw (ABJ) anomaly, allowing a second Cheshire coupling $\hat{\theta}_{B+L} \in \ZZ_3$.  

\begin{figure}[h]
    \centering
    \includegraphics[width=0.2\textwidth]{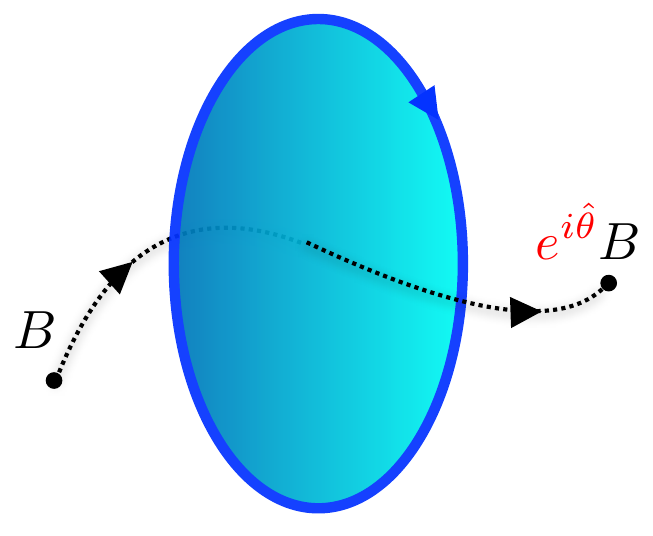}
    \caption{When quarks are massive, unit-winding axion strings (blue curve) are
    attached to finite-tension domain walls (light blue membrane).  The Cheshire $\hat{\theta}$ couplings produces an AB effect for baryons, so that they pick up a phase $e^{i\hat{\theta}}$ moving along the dotted black path. 
    }
    \label{fig:4d_AB_effect}
\end{figure}

The $2\pi$ periodicity of $\hat{\theta}$ in Eq.~\eqref{eq:theta_hat_QCD} follows from the general discussion in the preceding section. It can also be illustrated by an examination of the leading-order terms in the EFT. If $\hat{\theta}  = 2\pi k, k\in \ZZ$, we can perform a
globally-well-defined vector-like field redefinition of the quark $q$ and lepton $L$ fields $q \to e^{-i k \varphi}q$, $L \to e^{+i k \varphi}L$.  The resulting shifts of the quark and lepton
kinetic terms cancel Eq.~\eqref{eq:theta_hat_QCD}, illustrating the $2\pi$ periodicity.

Discrete symmetries do not force $\hat{\theta} = 0$ or $\hat{\theta}_{B+L} = 0$. The axion $\varphi$ is even under charge conjugation C and odd under parity P, so that $\hat{\theta}$ preserves CP but violates C and P.  The $2\pi$ periodicity of
$\hat{\theta}$ then implies that while $\hat{\theta} = 0$ is consistent with e.g. C symmetry, so is $\hat{\theta} = \pi$. But is C and P are anyway violated in the SM, so all values of $\hat{\theta} \in [0,2\pi)$ and $\hat{\theta}_{B+L} \in \ZZ_3$ are allowed by the symmetries of the axion-SM EFT.  Standard EFT principles then tell us that without fine-tuning in the UV completion, the Cheshire couplings $\hat{\theta}, \hat{\theta}_{B+L}$ should be $\mathcal{O}(1)$.

The $\hat{\theta}$ and $\hat{\theta}_{B+L}$ terms in the axion-SM EFT do not
affect the physics on $\RR^4$ in the absence of axion strings, so they
do not contribute to e.g. the neutron EDM.  But these Cheshire couplings do have a dramatic physical effect: they lead to AB-like phases.\footnote{A similar effect was found for photons passing by axion strings due to the axion-photon coupling~\cite{Harari:1992ea,Lue:1998mq,Pospelov:2008gg,Agrawal:2019lkr,Yin:2023vit}.}  For example, SM leptons and baryons\footnote{In the chiral effective field theory that describes the long-distance physics of QCD and axions, the baryon current maps to the tautologically-conserved Skyrme current, and the $\hat{\theta}$ term in Eq.~\eqref{eq:theta_hat_QCD} maps to a relatively standard example of a sigma-model $\theta$ term.} pick up AB phases $e^{i\hat{\theta}}$ when going around unit-winding axion strings.  This AB effect remains present even when the quark masses are non-zero, as illustrated in Fig.~\ref{fig:4d_AB_effect}.  

Finally, we note that in addition to the QCD axion, other axion-like particles (ALPs) are very common in UV completions of the SM, see e.g.~Refs.~\cite{Jaeckel:2010ni,Graham:2015ouw,Choi:2020rgn}. From the EFT perspective, all of them can have Cheshire $\hat{\theta}$ couplings to SM fields. Without fine-tuning,  generic axion strings should therefore have AB interactions with the SM fermions.

\vspace{0.5em}
{\bf Observational signature.} It is well known~\cite{Hindmarsh:1994re} that cosmic strings that carry a gauge flux $\Phi$ (in units of the fundamental flux quantum) generate Aharonov-Bohm-like phases $e^{i q \Phi}$~\cite{Aharonov:1959fk} for particles with gauge charge $q$.  This leads to the low-energy scattering cross-section for unit-charge fermions on a straight string~\cite{Alford:1988sj}
\begin{align}
\frac{d\sigma}{d \alpha}
=
\frac{\sin^{2}(q\Phi/2)}{2\pi p\sin^{2}\alpha/2} \,,
\label{eq:AB_sigma}
\end{align}
where $p$ and $\alpha$ are the momentum of the fermion and the scattering angle on the transverse plane, respectively.  
The $\hat{\theta}$ parameter makes SM fermions with $B-L$ charge $q_{B-L}$ moving around axion strings (associated with generic axion-like fields) pick up AB phases $e^{i q_{B-L} \hat{\theta}}$, so that their cross-sections also follow Eq.~\eqref{eq:AB_sigma}.  In contrast to other possible contributions to particle-on-axion-string scattering, this cross section is independent of all microscopic details of the axion model, and it is not suppressed by the e.g. $(p/f_{\rm PQ})^2$ or any perturbative coupling constants. 

\begin{figure}[t]
    \centering
    \includegraphics[width=0.48\textwidth]{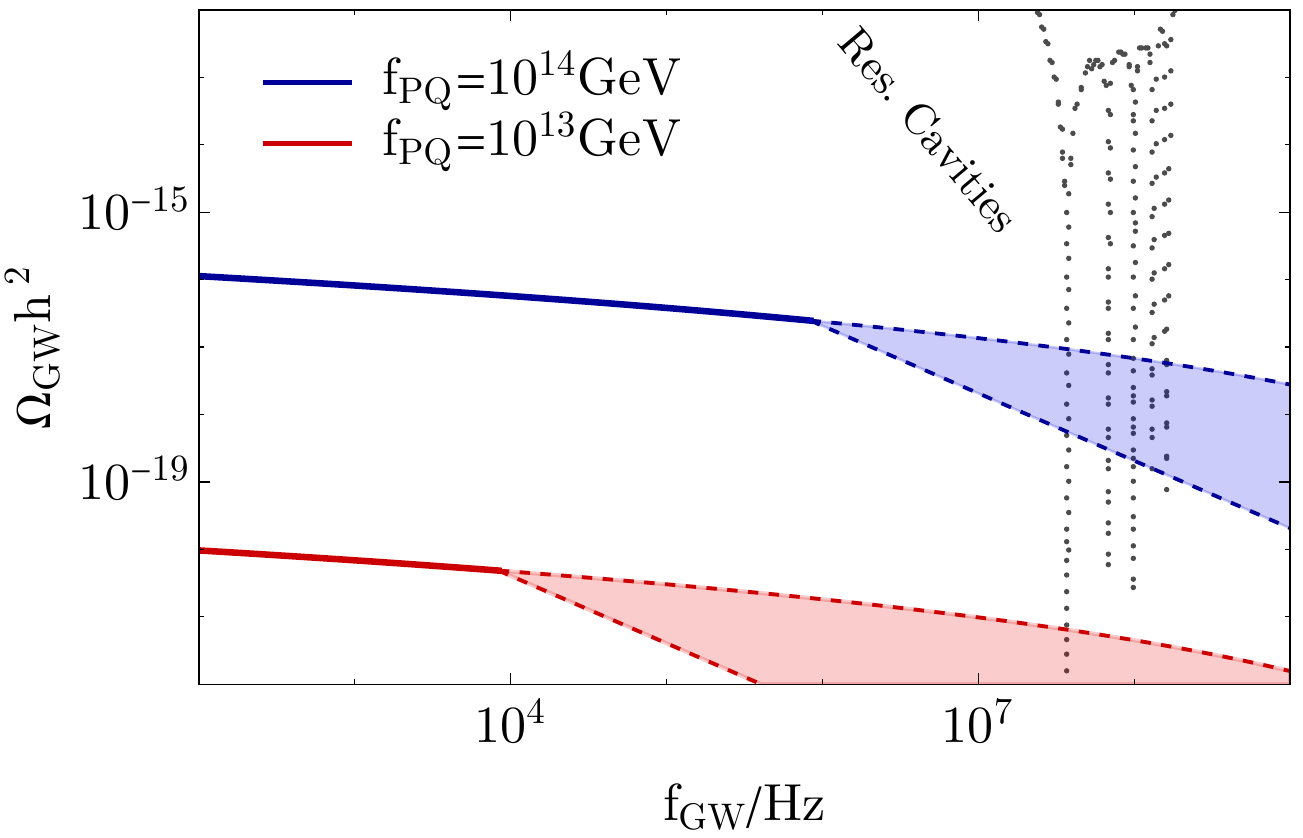}
    \caption{GW spectra from axion string networks for $f_{\rm PQ}=10^{14}{\rm GeV}$ (blue) and $10^{13}{\rm GeV}$ (red), with solid lines representing GW spectra with friction cutoffs at $f_{\rm fric}$ where the shaded regions begin. Without the Cheshire $\hat{\theta}$ couplings, the GW spectra would continue along the upper boundaries of the shaded regions, but $\hat{\theta}\neq 0$ induces drag against the thermal plasma, suppressing the GW spectra to lie somewhere within the shaded regions beyond $f_{\rm fric}$. The black dotted curve is the sensitivity of proposed resonant cavity detectors~\cite{Herman:2020wao,Herman:2022fau}. 
    }
    \label{fig:abgw}
\end{figure}

It is known~\cite{Vilenkin:1981kz,Vilenkin:1991zk} that cross-sections of the form Eq.~\eqref{eq:AB_sigma} lead to a drag force per unit length on cosmic strings $|\vec{f}_{\rm drag}|=\beta T^{3}\bar{v}$, where $\beta=(3\zeta(3)/4\pi^{2})\sum_{j}g_{*,j}\sin^{2}(\theta_{j}/2)$ is a drag coefficient with $g_{*,j}$ the number of relativistic degrees of freedom of a species $j$ with AB-like phase $\theta_j$, $T$ is the temperature, and $\bar{v}$ is a Lorentz-boosted relative velocity. Unless $\hat{\theta}$ is tuned to zero, axion strings will experience the same type of friction force as flux-carrying cosmic strings.  

Axion strings will be strongly affected by friction against the plasma of SM particles until either the particle number density becomes Boltzmann-suppressed or the background expansion overcomes the plasma friction.  The condition for this second effect is
$H>\ell_{\rm fric}^{-1}=\beta T^{3}/\mu$, where $\ell_{\rm fric}$ is a length scale for efficient friction and $\mu$ is the effective axion string tension $\mu\simeq2\pi f_{\rm PQ}^{2}\log(f_{\rm PQ}/H)$~\cite{Vilenkin:1981kz,Vilenkin:1991zk,Long:2014lxa}. During the radiation-dominated era, the temperature and Hubble expansion rate $H$ are related by $H = \sqrt{\pi^2 g_{*}/90} \,T^2/M_p$, where the Planck mass $M_p = 2.4 \times 10^{18}\,\textrm{GeV}$ and $g_{*}$ is the total effective number of relativistic degrees of freedom.  As a result, an axion string experiences a drag force due to quarks and leptons for temperatures greater than $T_{\rm fric}\simeq10^{11-2\delta}{\rm GeV}$ for $f_{\rm PQ} \simeq 10^{14-\delta}{\rm GeV}$.  Accordingly, when $\hat{\theta} = \mathcal{O}(1)$, axion string loop formation and oscillation gets suppressed 
for $T>T_{\rm fric}$. 

This plasma friction can affect the spectrum of gravitational waves (GWs) emitted from axion strings. The axion-string-induced stochastic gravitational wave background (SGWB) is mainly contributed by GWs emitted from long strings rather than small-scale structures including kinks and cusps~\cite{Krauss:1991qu,Jones-Smith:2007hib,Fenu:2009qf,Figueroa:2012kw}.
The present-day frequency of gravitational waves emitted from axion string loops formed at a plasma temperature $T$ is known to be given by $f(T)\simeq4.7\times10^{-6}{\rm Hz}(T/{\rm GeV})(g_{*}(T)/g_{*}(T_{0}))^{1/4}$~\cite{Gouttenoire:2019kij}, where $T_{0}=2.3\times10^{-4}{\rm eV}$ is the cosmic microwave background temperature.  
The plasma friction on axion strings coming from the Cheshire $\hat{\theta}$ term will introduce a UV-cutoff in the GW spectrum at $f_{\rm fric}\equiv f(T_{\rm fric})$ when compared to usual PQ scenarios like KSVZ~\cite{Kim:1979if,Shifman:1979if} and DFSZ~\cite{Zhitnitsky:1980tq,Dine:1981rt}. 

Current and near-term GW detectors have a chance of observing SGWB contributions from axions if $f_{\rm PQ}\gtrsim10^{14}{\rm GeV}$~\cite{Gorghetto:2021fsn,Gouttenoire:2019kij}, but the Hubble expansion rate during inflation has been constrained to $H_{\rm inf}/2\pi\lesssim9.6\times10^{12}{\rm GeV}$~\cite{Planck:2018jri}.  This naively seems to exclude post-inflation PQ breaking producing an observable gravity-wave signal from axion strings. But there are many mechanisms that can restore PQ symmetry after inflation ends, see e.g. Refs.~\cite{Kirzhnits:1972ut,Weinberg:1974hy,Kirzhnits:1974as,Kofman:1986wm,Vishniac:1986sk,Yokoyama:1989pa,Hodges:1991xs}, since the PQ scalar can get effective mass larger than its zero-temperature mass before reheating ends. 
If the axion strings of interest originate from the QCD axion, there can also be constraints on $f_{\rm PQ}$ from the relic abundance of dark matter.  These constraints are model-dependent, and there are axion models where QCD axions with $f_{\rm PQ} \sim 10^{14}\, \textrm{GeV}$ are not ruled out, see e.g. Refs.~\cite{Papageorgiou:2022prc,Choi:2022nlt,Agrawal:2017eqm,Kitajima:2017peg}.

Assuming a standard cosmology, we show an illustration of a sample GW spectrum from axion strings in Fig.~\ref{fig:abgw} assuming $\hat{\theta} = \mathcal{O}(1)$, with  solid curves that follow Eq. 165 of Ref.~\cite{Gouttenoire:2019kij}
up to a ``UV'' cutoff at $f = f_{\rm fric}$.
Beyond $f_{\rm fric}$, friction against the thermal plasma of SM fermions suppresses the GW spectrum to lie somewhere in the shaded regions. The lower dashed lines correspond to suppressed GW spectra emitted from the fundamental oscillation mode, while the upper edges show the unsuppressed spectra without the thermal friction. We leave a detailed determination of the GW spectra in the friction-suppressed region to future work. In any case, the suppression of the GW spectrum beyond $f\gtrsim f_{\rm fric}$ is an observational signature that might constrain the value of the Cheshire $\hat{\theta}$ terms discussed in this paper.

\vspace{0.5em}
{\bf Outlook.} We have explored some consequences of a novel `Cheshire' periodic coupling $\hat{\theta}$ in EFTs with a compact scalar field $\varphi$, as well as an Abelian global symmetry $G$ that does not act on $\varphi$. 
Such EFTs often appear as descriptions of phases where $\varphi$ is a (pseudo) Nambu-Goldstone boson of a spontaneously broken $0$-form $U(1)$ symmetry.   The Cheshire $\hat{\theta}$ coupling leads to AB effect for particles that are charged under $G$, so that they pick up phase shifts like $e^{i\hat{\theta}}$ as they move around $\varphi$ vortices.  This AB effect is `protected' by the $(D-2)$-form $U(1)$ vortex symmetry, and it is robust against explicit breaking of shift symmetries for $\varphi$.  We chose to illustrate the physical impact of Cheshire $\hat{\theta}$ terms by showing that they may have an observable impact on the physics of axions in particle physics. However, our observations can be applied to many other physical systems. 

Some questions opened by our results include:
\begin{itemize}
    \item The $\hat{\theta}$ term in Eq.~\eqref{eq:new_theta_generic} might arise from a current-current coupling $\frac{1}{\Lambda^2_{\rm UV}} \int d^{4}x j^{\mu}_{\mathcal{S}} j_{\mu}$ where $j_{\mathcal{S}}^{\mu}$ is the UV completion of the $\varphi$ shift symmetry current.  Are there well-motivated short-distance completions of such current-current interactions in e.g. the context of axion models? Are there other ways to induce $\hat{\theta}$?  
    
    \item Does the long-distance effective action for large $N$ QCD include a $\hat{\theta} = \pi$ term involving the Skyrme current and the $\eta'$ meson?
    
    \item AB-like phases for $U(1)$ global vortices have been discussed in many different contexts in e.g.~Refs.~\cite{Alford:1988sj,March-Russell:1991koo,Davis:1993jz,Chatterjee:2015lbf,Chatterjee:2019zwx,Alford:2018mqj,Cherman:2018jir,Cherman:2020hbe,Cherman:2021jmj,Hayashi:2023sas,Cherman:2024exo}.  Can some of them be related to our EFT parameter $\hat{\theta}$?
    \item In some situations, Cheshire $\hat{\theta}$ parameters become quantized.  Can EFTs with distinct quantized values of $\hat{\theta}$ parameters describe distinct phases of matter?
\end{itemize}
We hope that some of these questions can be answered in the future.

\vspace{0.5em}
{\bf Acknowledgements.} We thank Haipeng An, Jason L. Evans, Tony Gherghetta, Maxim Pospelov, Yuya Tanizaki, Arkady Vainshtein, and Uwe-Jens Wiese for helpful comments.  We are especially grateful to Seth Koren for suggesting that we think about friction on axion strings, and to Edward Hardy, Zohar Komargodski, Zhen Liu, and Tony Gherghetta for comments on a draft.  This work was supported by the Simons Foundation award number 994302 (A. C., S. C.), by the National Science Foundation Graduate
Research Fellowship under Grant No. 1842400 (M. N.), and in part by DOE grant DE-SC0011842 at the University of Minnesota (G. C.).

\bibliography{non_inv}
\end{document}